\documentclass{ws-rv9x6}
\usepackage{subfigure}   
\usepackage{ws-rv-thm}   
\usepackage{ws-rv-van}   
\usepackage{amsmath}
\usepackage{lineno}
\makeindex
\begin{document}

\chapter*{Open heavy flavour and quarkonium production as a function of the multiplicity in pp and p--Pb collisions with ALICE at the LHC}

\author[L. Valencia Palomo]{L. Valencia Palomo\footnote{For the ALICE collaboration.}}

\address{Benem\'{e}rita Universidad Aut\'{o}noma de Puebla\\
and Universidad Aut\'{o}noma de Chiapas, \\
lizardo.valencia.palomo@cern.ch}

\begin{abstract}
Due to the large masses of beauty and charm quarks,  their production cross sections can be computed in the framework of perturbative Quantum Chromodynamics. The correlation of quarkonium and open heavy-flavour hadron yields with charged particles produced in proton-proton (pp) and proton-lead (p-Pb) collisions can shed light on the interplay between hard and soft mechanisms in particle production. In this proceeding the results from D-meson and J/$\psi$ yields as a function of the charged-particle multiplicity in pp and p-Pb collisions are presented. Comparisons to theoretical models are also discussed.
\end{abstract}
\body

\section{Introduction}\label{Intro}

Quarkonium and open heavy-flavour production is still a challenging subject, as this is sensitive to both perturbative and non-perturbative aspects of Quantum Chromodynamics (QCD). The differential cross sections of heavy-flavour hadron production is computed using the factorization approach. In pp collisions, cross sections are calculated as a convolution of three terms: the Parton Distribution Functions (PDF) of the incoming nucleons, the partonic hard scattering cross section and the fragmentation function \cite{AndronicHFandOnium}.

At LHC energies, the measurement of heavy-quark production in the low transverse momentum ($p_{\mathrm{T}}$) region probes small values of the Bjorken-$x$ and large squared momentum transfers $Q^2$. Also, in this kinematic regime heavy-quark pairs are mainly created via gluon-gluon processes.

At high energies, pp collisions can have an important contribution from Multiple Partonic Interactions (MPI), where several interactions at the parton level can occur in a single pp collision \cite{Sjostrand1987}. The measurement of quarkonium and open heavy-flavour production in pp collisions as a function of the charged-particle multiplicity can provide insight into the processes occurring in the collision at the partonic level and the interplay between the hard and soft mechanisms in particle production. These aspects are expected to depend on the energy and on the impact parameter (the distance between the colliding protons in the plane perpendicular to the beam direction) of the pp collision \cite{ImpactParampp}.

Furthermore, measurements in pp collisions are used as reference for proton-nucleus (p-A) and nucleus-nucleus (A-A) studies. Characterizing the hot and dense medium produced in A-A collisions requires a quantitative understanding of the effects induced by the presence of nuclei in the initial state, the so-called Cold Nuclear Matter (CNM) effects. At the LHC these CNM effects are studied with p-Pb collisions.

Additional insight into CNM effects can be obtained by measuring the heavy-flavour hadron yields as a function of the multiplicity of charged particles produced in the p-Pb collision. The aim of these studies is to explore the dependence of heavy-flavour production on the collision geometry and on the density of final-state particles. Indeed, it is expected that the multiplicity of produced particles depends on the number of nucleons overlapping in the collision region.

\section{The ALICE detector}

Among the four main detectors at the LHC, ALICE is the only one that was designed and built to focus on high-energy heavy-ion collisions \cite{ALICE}. However, its programme also includes studies on pp and p-A collisions.

The ALICE detector can be divided in two parts: the central barrel and the muon arm. The central barrel is located at mid-rapidity and it allows studies of quarkonium decaying into two electrons but also open heavy flavours in their hadronic decays (D-mesons) and into single electrons. The muon arm is located at forward rapidity and it allows studies of quarkonium in the dimuon decay channel but also open heavy flavours via single muons.

The multiplicity estimations are provided by the number of track segments measured by the Silicon Pixel Detector (SPD) and the amplitude of the V0 scintillator. Due to their acceptance, SPD and V0 are the mid- and forward rapidity multiplicity estimators, respectively.

\section{Results}

In this section the results are presented as the J/$\psi$ or D-meson production relative yield (d$N_{\mathrm{J}/\psi,\mathrm{D}}$/d$y$)/$<$d$N_{\mathrm{J}/\psi,\mathrm{D}}\mathrm{/d}y>$ as a function of the relative charged particle multiplicity density (d$N_{\mathrm{Ch}}$/d$\eta$)/$<$d$N_{\mathrm{Ch}}\mathrm{/d}y>$ as determined with the mid- or forward rapidity multiplicity estimator.

\begin{figure}[h!]
\begin{center}
\begin{tabular}{cc} 
	\includegraphics[width=0.48\linewidth,height=5.8cm]{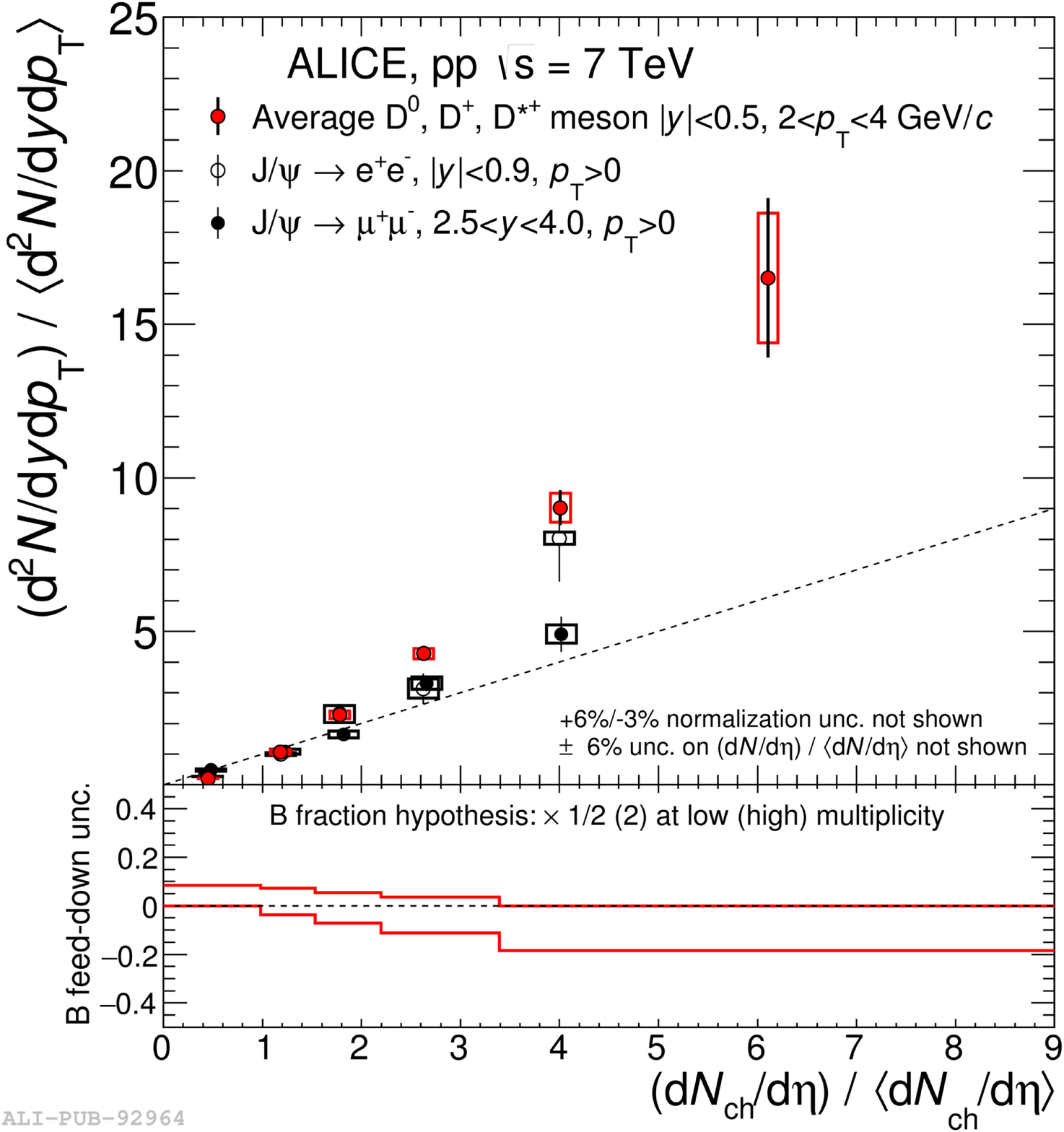} & 
   	\includegraphics[width=0.48\linewidth,height=5.8cm]{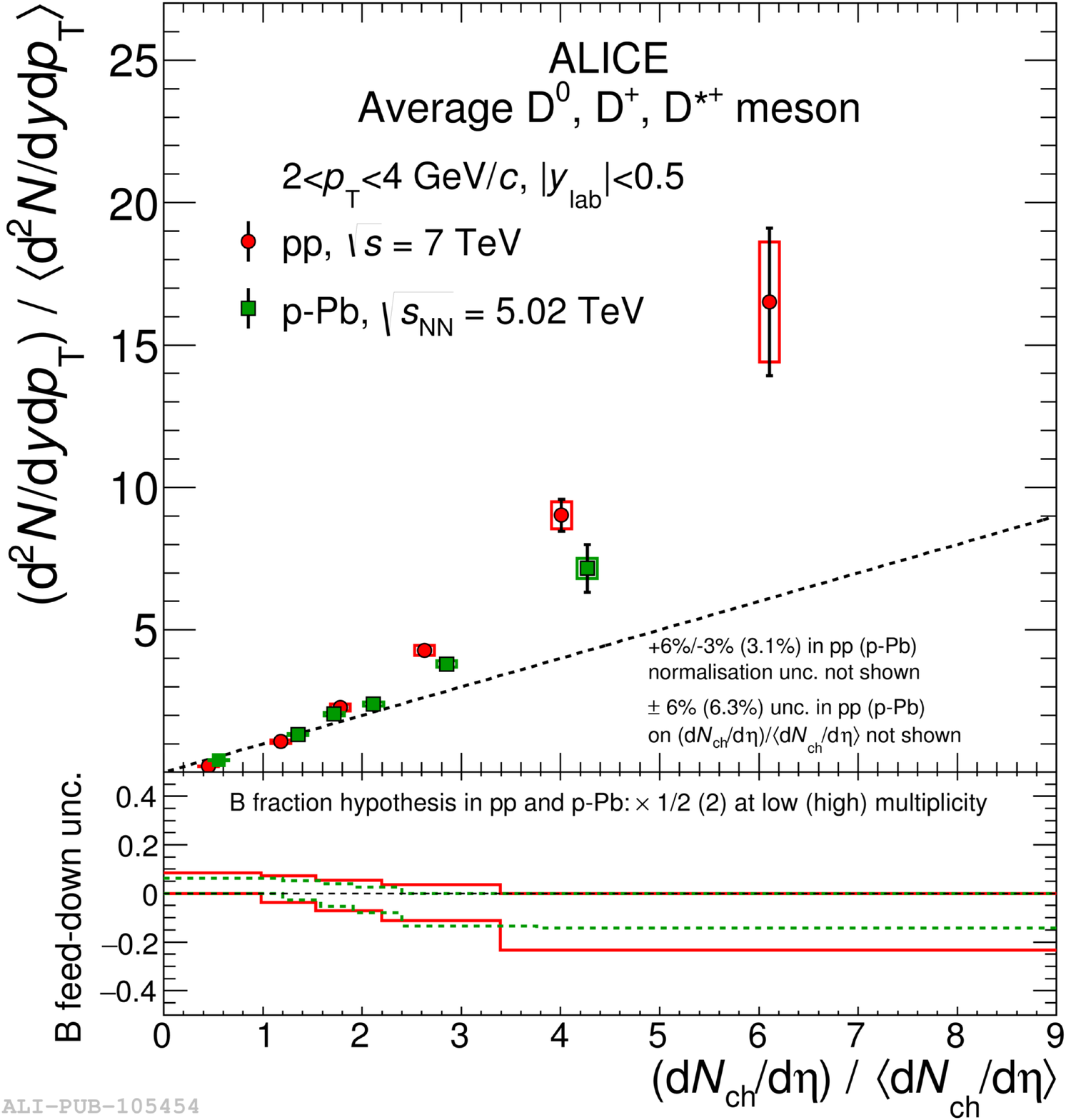} \\
\end{tabular}
\caption[aaaa]{\textit{Left:} Average D$^0$, D$^+$ and D$^{\ast +}$ relative yields in pp collisions at $\sqrt{s}$ = 7 TeV as a function of the relative charged-particle multiplicity compared to J/$\psi$ \citep{CandBinpp}. \textit{Right:} D-mesons results in p-Pb collisions at $\sqrt{s_{\mathrm{NN}}}$ = 5.02 TeV compared to pp. In both plots the SPD is used as multiplicity estimator \citep{DinpPb}.}
\label{FigExp}
\end{center}
\end{figure}

The result of the average prompt D$^0$, D$^+$ and D$^{\ast +}$ mesons relative yields for $2 < p_{\mathrm{T}} < 4$ GeV/$c$ in pp collisions at $\sqrt{s}$ = 7 TeV as a function of the relative charged-particle multiplicity is shown on the left side of figure \ref{FigExp} \citep{CandBinpp}. This plot also contains the results for inclusive J/$\psi$ measured at mid-rapidity (J/$\psi$ $\rightarrow$ e$^-$ + e$^+$) and at forward rapidity (J/$\psi$ $\rightarrow$ $\mu^-$ + $\mu^+$) \citep{Jpsi2012}. The multiplicity estimator used was the mid-rapidity one, statistical errors are shown as vertical bars and systemtic uncertainties appear as open boxes. For D-mesons the uncertainty on the feed-down fraction is shown at the bottom of the plot. Both for D- and J/$\psi$ mesons there is a faster than linear increase and for the D-mesons it is observed that this behaviour is independent of $p_{\mathrm{T}}$.

In ALICE, prompt and non-prompt J/$\psi$ components can only be separated at mid-rapidity. At forward rapidity only inclusive J/$\psi$ can be measured. Prompt J/$\psi$ yields have a similar result as the inclusive one, while for the non-prompt component the large statistical errors avoid to draw a firm conclusion \citep{CandBinpp}.

A faster than linear increase is also observed for D-mesons using the forward rapidity multiplicity estimator \citep{CandBinpp}. The idea behind using V0 to measure the multiplicity, is to minimise the influence of particles produced in the charm fragmentation and D-meson decay in the multiplicity estimation.

The right plot of figure \ref{FigExp} shows the comparison between D-meson measurements in p-Pb collisions at $\sqrt{s_{\mathrm{NN}}}$ = 5.02 TeV and the pp ones, both with the mid-rapidity multiplicity estimator \citep{DinpPb}. A faster than linear increase is also observed in p-Pb collisions, however the increase in pp seems to be faster. The behaviour observed in p-Pb events is also independent of $p_{\mathrm{T}}$.

\begin{figure}[h!]
\begin{center}
\begin{tabular}{c} 
\includegraphics[width=10.0cm,height=9cm]{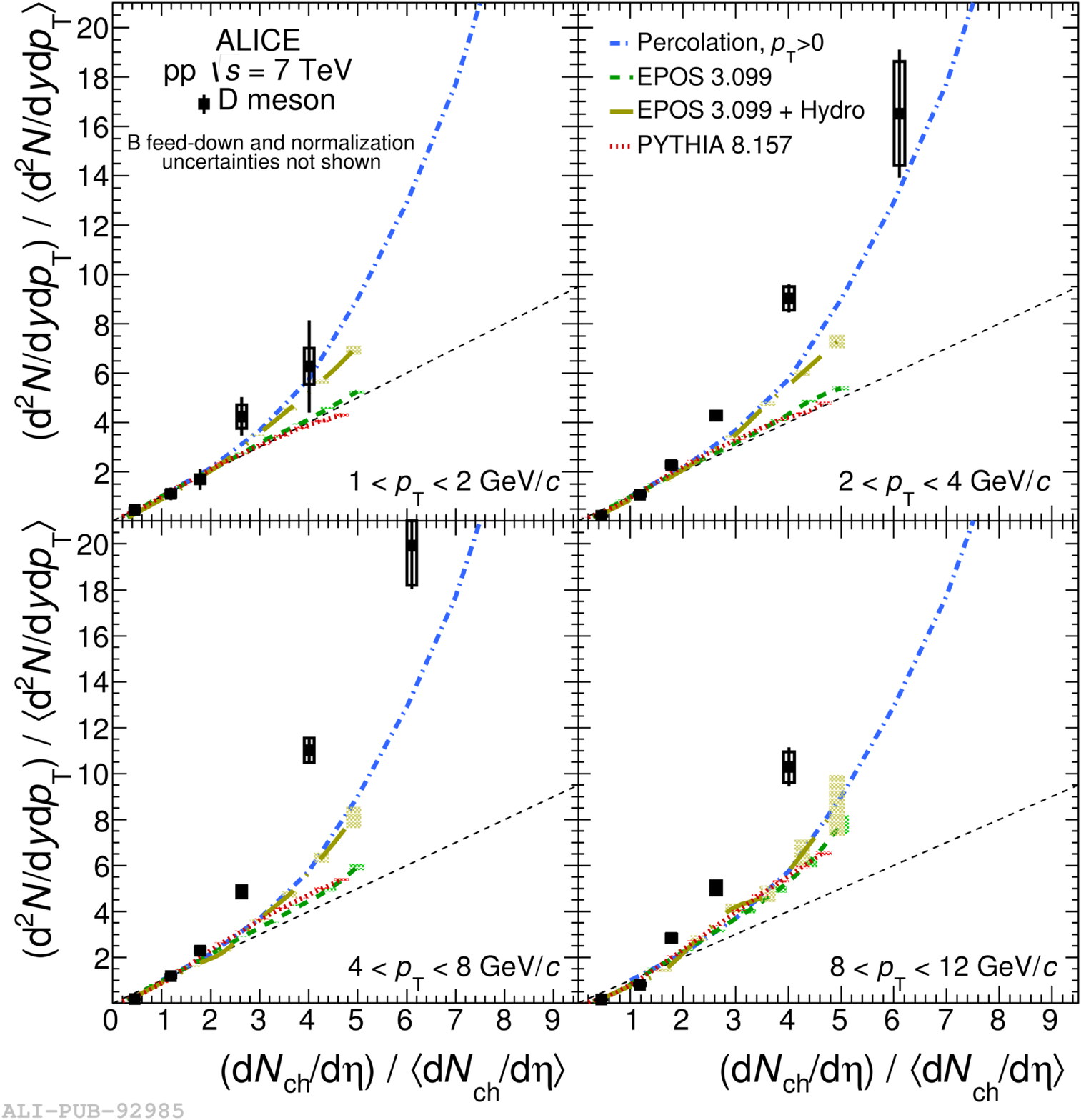}
\end{tabular}
\caption[aaaa]{D-mesons results in pp collisions compared to theoretical models \citep{CandBinpp}.}
\label{FigTheoryI}
\end{center}
\end{figure}

Figure \ref{FigTheoryI} presents the D-meson results in pp collisions compared to theoretical models:
\begin{itemize}
\item Pythia 8 with $``$SoftQCD$"$ process selection includes colour reconnection and diffractive processes. Main open heavy-flavour contibution comes from Initial and Final State Radiation followed by hard processes in MPI.
\item EPOS 3 imposes the same theoretical framework for various colliding systems: pp, p-A and A-A. Initial conditions generated in the Gribov-Regge multiple scattering framework followed by a hydrodynamical evolution. Hadronisation is performed with a string fragmentation procedure. Two kinds of predictions are available: with and without the hydrodynamical evolution.
\item Percolation model assumes the exchange of colour sources between the projectile and the target, these colour sources have a finite spatial extension and can interact. The prediction for D-mesons shown in all panels of the figure is $p_{\mathrm{T}}$-integrated. 
\end{itemize} 

For percolation and EPOS + hydro, the increase with multiplicity of the model predictions deviates from the linear behaviour as the $p_{\mathrm{T}}$ increases. EPOS without hydro and Pythia calculations are always linear, but the slope increases more than one as the transverse momentum becomes larger. Except for the lowest $p_{\mathrm{T}}$ bin, the measured data has an even faster increase than the model predictions. However, the experimental results tend to favour calculations with a substantial deviation from linearity at high multiplicities such as EPOS 3 with hydrodynamics or the percolation model.

\begin{figure}[h!]
\begin{center}
\begin{tabular}{c} 
\includegraphics[width=10.0cm,height=9cm]{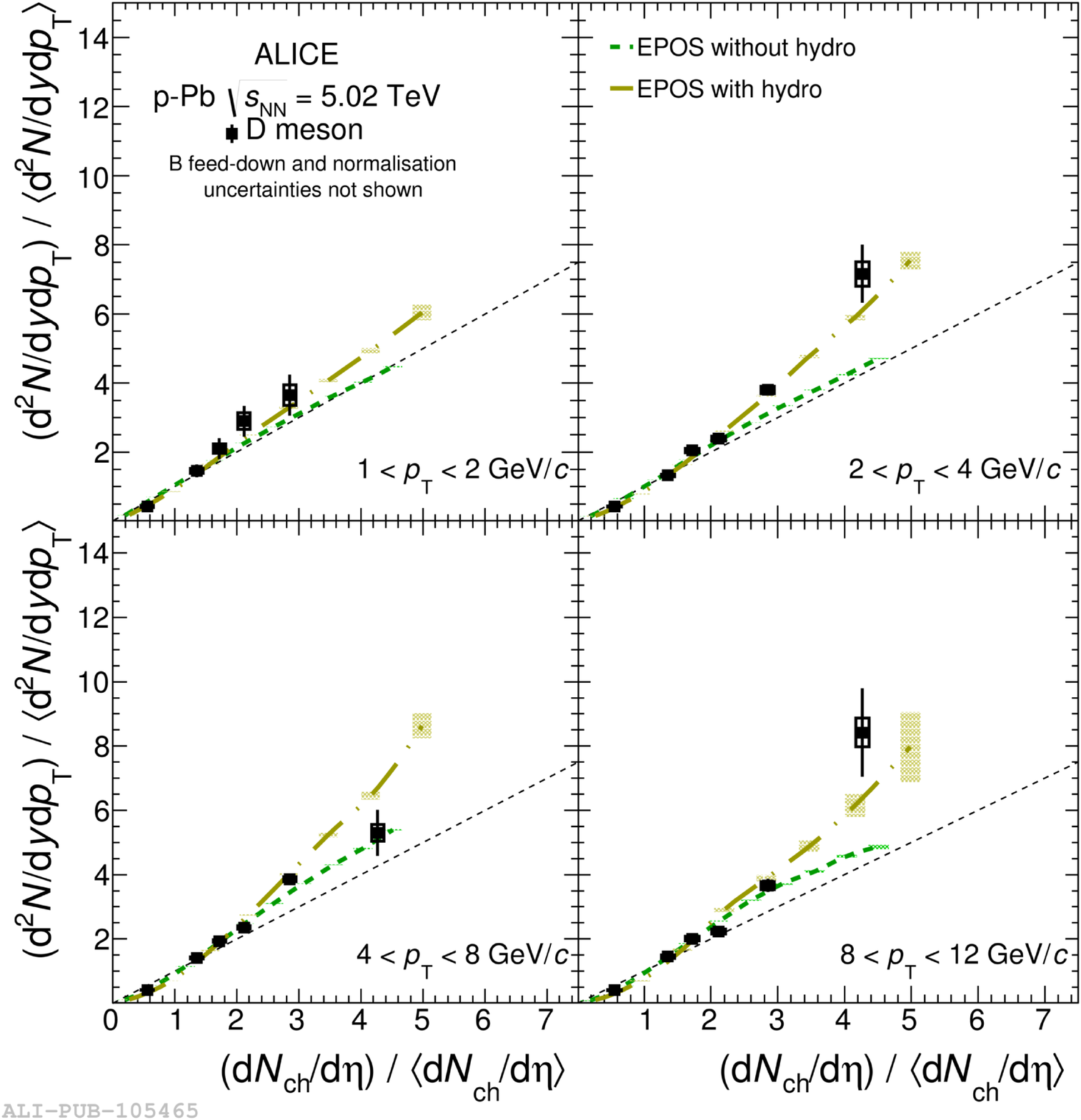}
\end{tabular}
\caption[aaaa]{D-mesons results in p-Pb collisions compared to EPOS 3 with and without hydrodynamical evolution \citep{DinpPb}.}
\label{FigTheoryII}
\end{center}
\end{figure}

Figure \ref{FigTheoryII} presents the D-meson results in p-Pb collisions compared to EPOS 3 with and without hydrodynamical evolution. The experimental results agree with the EPOS 3 model calculations within uncertainties. Results at high multiplicity are better reproduced by the calculation including a viscous hydrodynamical evolution of the collision, which predicts a faster-than-linear increase of the charmed-meson yield with multiplicity.

\section{Conclusions}

Quarkonium and open heavy-flavour production as a function of the multiplicity in pp and p-Pb collisions are useful tests for the Multiple Parton Interactions scenario. In pp collisions there is a faster than linear increase both for J/$\psi$ and D-mesons, an indication that there is small influence of hadronisation. Models including MPI can reproduce the data. In p-Pb collisions there is a faster than linear increase at high multiplicities, but this is slower than in pp. In this case, results can be described by the EPOS with hydro 3 event generator. 

\bibliographystyle{ws-rv-van}
\bibliography{ms}

\end{document}